 \def\Title{Error in an argument regarding ``improper'' mixtures}
 \def\arXiv{quant-ph/0405058}
 \def\Abstract{%
An argument, perhaps originating with Feyerabend a half century ago, and repeated many times since, purporting to establish that an ``ignorance interpretation'' of a bipartite pure entangled state leads to logical inconsistency, is incorrect: the argument fails to account for the effects of indistinguishability. 
}%
\documentclass[aps,rmp,nobibnotes,nofootinbib,onecolumn,twoside,noshowpacs,draft]{revtex4}
 \usepackage[tbtags,sumlimits]{amsmath}
 \usepackage{amssymb}

 \makeatletter
 \def\p@section{}
 \def\p@subsection{}
 \def\p@subsubsection{}
 \def\p@paragraph{}
 \def\p@subparagraph{}

 \def\@hangfrom@section#1#2#3{\@hangfrom{#1}{\large\textrm{#2}}{\large\textrm{#3}}}%
 \def\@hangfrom@subsection#1#2#3{\@hangfrom{#1}{\textrm{#2}}{\textrm{#3}}}%
 \def\@hangfrom@subsubsection#1#2#3{\@hangfrom{#1}{\textrm{#2}}{\textrm{#3}}}%
 \def\frontmatter@setup{\normalfont\raggedright}
 \makeatother

 \usepackage{xspace}

 \newcommand{\eg}{e.g., }

 \newcommand{\dEsp}{d'Espagnat\xspace}

 \newcommand{\set}[1]{\ensuremath{{\left\{\,#1\,\right\}}}}%

 \newcommand{\Sys}[1]{\ensuremath{\mathcal{#1}}\xspace}

 \newcommand{\ket}[1]{\ensuremath{\vert\,{#1}\,\rangle}}
 \newcommand{\bra}[1]{\ensuremath{\langle\,#1\,\vert}}
 \newcommand{\proj}[1]{\ensuremath{\ket{{#1}}\bra{{#1}}}}

 \newcommand{\PSI}[2]{\Psi^{\Sys{#1}}_{\,#2}}%
 \newcommand{\ketPsi}[2]{\ensuremath{\left|\PSI{#1}{#2}\right\rangle}\xspace}

 \newcommand{\bRho}{\pmb{\rho}}
 \newcommand{\Rho}[2]{\ensuremath{\bRho^{\Sys{#1}}_{#2}}\xspace}

 \newcommand{\Trace}[2][]{\ensuremath{{\rm Tr}%
                    {\!}_{\Sys{#1}}\left\{\,{#2}\right\}}}

 \newcommand{\SysA}{\Sys{A}}
 \newcommand{\SysB}{\Sys{B}}
 \newcommand{\SysAB}{\Sys{A\oplus B}}

 \newcommand{\ketPsiAB}{\ketPsi{AB}{}}
 \newcommand{\RhoA}[1][]{\Rho{A}{#1}}
 \newcommand{\RhoB}[1][]{\Rho{B}{#1}}
 \newcommand{\RhoAB}[1][]{\Rho{A\oplus B}{#1}}

\begin{document}
 \makeatletter
 \def\ps@titlepage{%
   \renewcommand{\@oddfoot}{}%
   \renewcommand{\@evenfoot}{}%
   \renewcommand{\@oddhead}{\hfill\arXiv}
   \renewcommand{\@evenhead}{}}
 \makeatother

\title[Kirkpatrick -- \Title] 
      {\Title} 

\author{K.~A.~Kirkpatrick}
\email[E-mail: ]{kirkpatrick@physics.nmhu.edu}
\affiliation{New Mexico Highlands University, Las Vegas, New Mexico 87701}
\begin{abstract}
 \Abstract
\end{abstract}
 \maketitle
 \makeatletter\markboth{\hfill\@shorttitle\hfill}{\hfill\@shorttitle\hfill}\makeatother
 \pagestyle{myheadings}

\section{Introduction}

An ``ignorance interpretation'' of probability%
\footnote{%
The term ``ignorance'' itself arises from the misconception that the probability is due to not knowing which element will appear next --- that if the sequence were known, all probabilities would be 0 or 1. To the extent that probability relates in any way to frequency of occurrence, this is incorrect: \eg dealing hands of poker, pairs will occur with the correct statistical frequency even if the shuffling is done with a predictable pseudo-random number generator.
} %
refers to random selection from already-existing elements (such as drawing cards from a deck). In quantum mechanics, the ignorance interpretation takes a specific form: The statistical operator $\bRho=\sum_t w_t\,\proj{\phi_t}$  represents a system which, with probability $w_j$, is ``really in'' one or another of the pure states \ket{\phi_j}. 

In quantum mechanics, mixtures are thought to arise in two ways --- as a direct mixing of distinct pure-state preparations, and ``improperly'' (\dEsp's term), as the reduction of a pure-state composite system to the state of a component system by tracing out the other component systems. An Argument has appeared many times in the past half-century to the effect that an attempt to apply an ignorance interpretation to an ``improper'' mixture leads to mathematical inconsistency. Specifically: The state of a composite system, assumed pure, is reconstructed from the purported ignorance-interpretation subensembles arising from the trace-reductions, and it is claimed that this reconstruction leads to a mixed, rather than pure, state. It is the purpose of this note to show that this Argument is incorrect. (It is \emph{not} the purpose of this paper to argue further regarding the possibility of such ignorance interpretations.)

\section{The Argument}\label{S:erroneous}%

Here is a simple presentation of the Argument:
\begin{enumerate}\item[]%
Consider a composite system \SysAB in the pure state $\ketPsiAB=\sum_t\,c_t\,\ket{\alpha_{t}\,\beta_{t}}$. The component states are the mixed states $\RhoA=\Trace[\,B]{\proj{\Psi^{AB}}}=\sum_t\,|c_t|^2\,\proj{\alpha_{t}}$ and $\RhoB=\Trace[\,A]{\proj{\Psi^{AB}}}=\sum_t\,|c_t|^2\,\proj{\beta_{t}}$. If an ``ignorance interpretation'' of these states holds --- if system \SysA is really in one of the pure states \set{\ket{\alpha_{j}}} (with probability $|c_j|^2$), and system \SysB is really in one of the pure states \set{\ket{\beta_{j}}} (with probability $|c_j|^2$) --- then, taking into account the correlations implied by \ketPsiAB, the composite system \SysAB is really in one of the pure states \set{\ket{\alpha_{j}\,\beta_{j}}} with probability $|c_j|^2$. Therefore, this Argument concludes, the  composite system is in the \emph{mixed} state $\rule{0pt}{9pt}\RhoAB=\sum_t|c_t|^2\,\proj{\alpha_{t}\,\beta_{t}}$, contradicting the assumption that \SysAB is in a pure state.
\end{enumerate}
This Argument is implicit in \citet[p.~124]{Feyerabend57}, is explicit in d'Espagnat (\eg \citeyear{dEspagnat76}, pp.~59-61), and appears frequently in the literature of quantum interpretation --- \eg \citet[p.~150 and p.~283]{Hughes89}, \citet[p.~207]{vanFraassen91book}, \citet{Peres94}, among many others. Clearly, it is widely accepted.

\section{The error}\label{S:error}

However, the contradiction claimed by the Argument is not supportable. The Argument assumes that any combination of states is a mixture; this parallels classical mixtures --- but in classical systems, there is no issue of indistinguishability. In the present case, however, nothing \emph{external} to \SysAB distinguishes the states \set{\ket{\alpha_{j}\,\beta_{j}}} from one another (we know this from the fact that \SysAB is in a \emph{pure} state), so by standard rules of quantum mechanics the state of the composite system must be a superposition of these \emph{indistinguishable} component states, a pure state. The claimed contradiction does not obtain --- the Argument fails to establish anything inconsistent about an ignorance interpretation of these mixtures.

Alternatively, we could construct the Argument's ensemble of states by combining preparations of \set{\ket{\alpha_{j}\,\beta_{j}}} randomly with the appropriate probabilities $|c_j|^2$.%
\footnote{%
This follows the latest presentation of the Argument by \citet[p.~104]{dEspagnat95}, who did not, however, consider the role of the indistinguishability of the preparations.
} %
If those preparations were indistinguishable (in the sense that, given an occurrence of the preparation, it is \emph{in principle} impossible to determine from which preparation apparatus it comes), the states would combine coherently,%
\footnote{%
This is well established experimentally, for example by numerous results of the Rochester group (from \citet{PfleegorMandel67} to \citet*{WangZouMandel91}).
} %
again resulting in no contradiction.

\section{Conclusion}%
It is established that this Argument fails, and thus can present no objection to the ignorance interpretation of mixtures arising from an entangled joint state. 

However, to establish an error in an argument against the ignorance interpretation is not to argue that the ignorance interpretation is correct. Nor does a presentation of the Argument, with its assumption of an ignorance interpretation (each occurrent system of the mixture being ``really in'' a pure state), require an explanation of exactly what ``to be in'' a state would mean, beyond what is needed for the statement of the Argument, nor is a defense of that ignorance interpretation in the face of \emph{other} objections and difficulties required: to the extent that the ignorance interpretation is inexplicable or indefensible, to that same extent the Argument fails before it can be stated.

 \renewcommand{\refname}{\sc References}
 \footnotesize%


\begin{thebibliography}{}

\bibitem[{d}'Espagnat(1976)]{dEspagnat76}
{d}'Espagnat, B. (1976).
{\em Conceptual Foundations of Quantum Mechanics}.
W.~A.~Benjamin, Menlo Park, CA, 2nd edition.

\bibitem[{d}'Espagnat(1995)]{dEspagnat95}
------ (1995).
{\em Veiled Reality}.
Addison-Wesley, Reading, MA.

\bibitem[Feyerabend(1957)]{Feyerabend57}
Feyerabend, P.~K. (1957).
On the quantum theory of measurement.
In K{\"o}rner, S., editor, 
{\em Observation and Interpretation: A Symposium  of Philosophers and Physicists}, 
pages 121--130. Butterworth, London. (Reprinted 1962 by Dover Publications, New York.)

\bibitem[Hughes(1989)]{Hughes89}
Hughes, R.~I.~G. (1989).
{\em The Structure and Interpretation of Quantum Mechanics}.
Harvard University Press, Cambridge, MA.

\bibitem[Peres(1994)]{Peres94}
Peres, A. (1994).
``Time asymmetry in quantum mechanics: A retrodiction paradox,'' 
\emph{Phys. Lett. A} {\bf 194}, 21--25.

\bibitem[Pfleegor and Mandel(1967)]{PfleegorMandel67}
Pfleegor, R.~L. and Mandel, L. (1967).
``Interference of independent photon beams,'' 
\emph{Phys. Rev.} {\bf 159}, 1084--1088.

\bibitem[{v}an Fraassen(1991)]{vanFraassen91book}
{v}an Fraassen, B.~C. (1991).
{\em Quantum Mechanics: {A}n Empiricist View}.
Oxford University Press.

\bibitem[Wang et~al.(1991)Wang, Zou, and Mandel]{WangZouMandel91}
Wang, L.~J., Zou, X.~Y., and Mandel, L. (1991).
``Induced coherence without induced emission,'' 
\emph{Phys. Rev. A} {\bf44}(7), 4614--4622.

\end{thebibliography}
\end{document}